\documentclass[sigconf]{acmart}

\usepackage{booktabs} 
\usepackage{multirow}
\usepackage{hhline}

\begin{document}
\title{Measuring Average Treatment Effect from Heavy-tailed Data}

\author{Jason (Xiao) Wang}
\affiliation{%
  \institution{eBay Inc.}
  \streetaddress{2025 Hamilton Avenue}
  \city{San Jose}
  \state{CA}
  \postcode{95125}
}
\email{xwang@ebay.com}

\author{Pauline Burke}
\affiliation{%
  \institution{eBay Inc.}
  \streetaddress{2025 Hamilton Avenue}
  \city{San Jose}
  \state{CA}
  \postcode{95125}
}
\email{pmburke@ebay.com}


\begin{abstract}
Heavy-tailed metrics are common and often critical to product evaluation in the online world. While we may have samples large enough for Central Limit Theorem to kick in, experimentation is challenging due to the wide confidence interval of estimation. We demonstrate the pressure by running A/A simulations with customer spending data from a large-scale Ecommerce site. Solutions are then explored. On one front we address the heavy tail directly and highlight the often ignored nuances of winsorization. In particular, the legitimacy of false positive rate could be at risk. We are further inspired by the idea of robust statistics and introduce Huber regression as a better way to measure treatment effect. On another front covariates from pre-experiment period are exploited. Although they are independent to assignment and potentially explain the variation of response well, concerns are that models are learned against prediction error rather than the bias of parameter. We find the framework of orthogonal learning useful, matching not raw observations but residuals from two predictions, one towards the response and the other towards the assignment. Robust regression is readily integrated, together with cross-fitting. The final design is proven highly effective in driving down variance at the same time controlling bias. It is empowering our daily practice and hopefull can also benefit other applications in the industry.
\end{abstract}

%
%
%

\keywords{A/B test, controlled experiment, heavy tail, robust statistics, Huber estimator, debiased machine learning, post-stratification}

\maketitle

\section{Introduction}
Controlled experiment or A/B testing is widely used to evaluate product innovations in Internet industry and companies like Microsoft, Google and LinkedIn have built sophisticated systems to meet their highly demanding needs \cite{Tang10,Kohavi13,Xu15}. At eBay we have seen continuous growth of the usage in past decade. Our in-house system now runs thousands of experiments a year, impacting tens of millions of customers on a daily basis across countries, channels and a variety of products \cite{Wang18}. What is typically employed is randomly selecting a group of customers to offer the new experience (\textit{treatment}), while giving another group the status quo (\textit{control}). Randomization helps protect inference against confounding factors. With Rubin's causal model \cite{Rubin90} the average treatment effect (ATE) can be estimated using the observed difference
\begin{equation}
\hat\mu=\bar Y_{i \in treatment} - \bar Y_{i \in control}
\label{eq1}
\end{equation} 
where $Y_i$ is the response metric from customer $i$. 

\begin{figure}
\centerline{\includegraphics[scale=0.38]{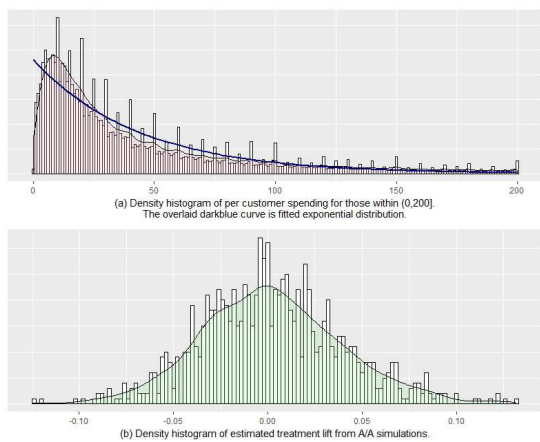}}
\caption{The heavy tail of per customer spending and its impact on inference.}
\label{fig1}
\end{figure}

Our system reports hundreds of metrics for each experiment, measuring a wide range of activities from pageviews, click-through to transactions. Most of them are not Gaussian distributed. Usually it is not a problem as with sample size in the millions, by Central Limit Theorem the distribution of sample mean can be closely approximated by Gaussian. A two-sample $z$ (or $t$)-test is justified. If the p-value turns out to be smaller than a pre-defined threshold (e.g. 0.1), null hypothesis is rejected and the treatment is declared having a \textit{statistically significant} impact. However, heavy-tailed distributions are frequently seen and pose a big challenge. Take the spending on bought merchandise as an example which is critical for commerce. We have a ten million sample aggregated over a two weeks period. $89\%$ of those converted customers (a small portion of the sample) have a spending less than \$200 while the top 1\% spend anywhere from \$1250 to a couple hundred thousand, contributing 28\% to the total spending\footnote{The numbers are not from current traffic and not representative to eBay's sales.}. Fig. \ref{fig1}(a) shows the distribution within (0, 200] range and it suggests the right tail is not bounded by the exponential (darkblue) curve. Distribution beyond the range is not amenable to visualization but obeys the pattern. 

Heavy tail implies that observations in the high percentile impose a large influence on sample mean. When the variance is near infinite, the convergence of sample mean may be frustratingly slow \cite{Athreya87}. To address the concern we run a thousand A/A simulations where a random subset of two million is picked to form the treatment while another subset of two million forms the control. ATE is estimated using (\ref{eq1}) and further normalized to produce the \textit{treatment lift}. Its distribution is shown in Fig. \ref{fig1}(b).   
While a normality test (e.g. Shapiro-Wilk) won't reject the Gaussian hypothesis, there is asymmetry around zero. More importantly, the standard deviation is estimated to be $0.0375$. Using a 95\% confidence interval ($+/-2\sigma$) it translates to detectability of minimum 7.5\% lift in average customer spending. Most of our experiments won't have such a big impact with nowaday's agile strategy. Larger sample size helps but is often unaffordable. The result is disappointing and even deemed impractical by our product teams. 

A direct answer is to limit the over-sized influence of extreme observations. Trimming goes too far in our view to completely remove those observations from analysis. Winsorization \cite{Dixon60} as a popular design replaces values above and/or below a threshold by that threshold. Its conceptual and computational simplicity is very attractive which is why winsorization becomes our first choice at eBay. Nevertheless, estimation is seen very sensitive to the threshold and there are no established guidelines for its optimal selection. People also look into bootstrap for inference with heavy-tailed data. Athreya \cite{Athreya87} confirms that conventional approach of resampling is not consistent if the variance does not exist. Despite some remedy, both \textit{m out of n} boostrap and subsampling can be quite unreliable \cite{Cornea15}. Taddy et al. \cite{Taddy16} build a semi-parametric process which combines resampling for the bulk of the distribution and Bayesian parametric model for the tail. Although it demonstrates low error rate as well as reliable uncertainty quantification, cost is a major concern. We run a large number of computations everyday and boostrap at scale doesn't gain any support from the team.

Another line of thought is to bring in covariates for variance reduction. Deng et al. \cite{Deng13} propose to use pre-experiment data in online experiments and report very promising results. Their model assumes the response is a linear addition of effects from treatment and covariates. Recently there is growing appeal to employ advanced machine learning (ML) algorithms in inference, replacing the linear model with those learned from regularized regression, random forest, neural net and others \cite{Athey18}. The problem is that high predictive power doesn't necessarily lead to consistent parameter estimation. Techniques like doubly robust \cite{Robins95} and double/debiased learning \cite{Cherno17} are hence developed. It is of great interest to see how they can help but we have yet to see any empirical study from the industry.

Our goal is to deliver highly reliable experiments in the presence of heavy tail and this paper documents the effort and contributions. Specifically,
\begin{itemize}
\item We dive into the intricacies of winsorization arised in the experimentation in section \ref{section2}. The risk of disproportionate false positives is revealed. Also studied are the optimization of threshold and its dependency on sample size and treatment effect. The investigation relies deeply upon a mixture model of the distribution of response metric and simulation. 
\item Section \ref{section3} is inspired by robust statistics and the M-estimators. We propose Huber regression as a novel design and compare its performance with winsorization.
\item Section \ref{section4} shifts attention to the incorporation of covariates and how they can be exploited in the best way. We don't go too far to concern heterogeneity but focus on applying ML techniques in measuring the average effect. We are able to demonstrate very good performance of a design combining double/debiased learning and Huber regression.
\item Section \ref{section5} concludes the paper and talks briefly about the future work.
\end{itemize}

\section{Winsorization in practice}
\label{section2}

Most of our metrics are right-tailed therefore winsorization, also known as capping, sets to replace $Y_i$ with $min(Y_i, a)$ and produce a capped estimation of treatment effect, $\hat\mu_a$. $a$ is the threshold and usually chosen to be a high percentile of the sample. To determine its optimal value Kokic and Bell \cite{Kokic94} go by minimizing the mean square error (MSE) of estimation against the true effect $\mu$ (derived from a stratified mean model):
\begin{equation}
\label{eq_capping}
\operatorname*{argmin}_a \mathop{\mathbb{E}}(\hat\mu_a-\mu)^2.
\end{equation}
A similar approach is taken here. Obviously we need know $\mu$ and also generate samples in correspondence, so as to produce $\hat\mu_a$. Fundamental to the ability is a good representation of the data generating process of $Y_i$.

Distributions like Fig. \ref{fig1}(a) are characterized by density spikes at discrete price points (e.g. \$0.99, \$4.99, etc.) which are not tractable by low-dimensional parametric families. Certain approximation error is inevitable. We have tried distributions of exponential, log-normal, log-weibull, etc. and eventually land on the idea of piecewise mixture model. Specs will be different for different metrics. For customer spending we choose to have three pieces: non-converting customers, torso of converted customers (spending below a cutoff) and the tail (spending beyond the cutoff). We assume the torso is realized from a truncated exponential distribution (with rate $\lambda$ and upper bound $C$) and the tail is realized from a type I Pareto (with shape $\alpha$ and scale $C$, same as the torso's upper bound). To put it in equation,
\begin{equation} 
\label{eq_dis}
P(Y)=\left\{
		\begin{array}{ll}
		p_{\text{non-conversion}}, & Y=0\\
		p_{\text{torso}}*\frac{\lambda}{1-e^{-\lambda C}}*e^{-\lambda Y}, & Y\in(0,C)\\
		p_{\text{tail}}*\frac{\alpha*C^\alpha}{Y^{\alpha+1}}, & Y\in[C,+\infty)
		\end{array}
	\right.
\end{equation}
provided $p_{\text{non-conversion}}+p_{\text{torso}}+p_{\text{tail}}=1$. Mean of $Y$ is then derived as
\begin{equation}
\label{eq_mean} 
\mathop{\mathbb{E}}(Y)=p_{\text{torso}}*(\frac{1}{\lambda}-\frac{C}{e^{\lambda C}-1}) + p_{\text{tail}}*\frac{\alpha C}{\alpha-1}.
\end{equation}

\begin{figure}
\centerline{\includegraphics[scale=0.41]{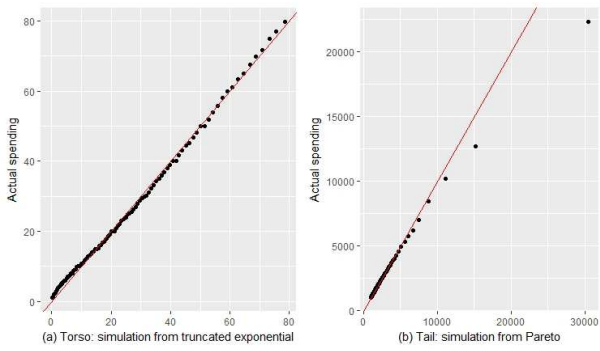}}
\caption{QQ plots of the torso and tail segments comparing simulation from estimated model (horizental) to actual spending (vertical).}
\label{fig2}
\end{figure}

Among the parameters $p_{\text{non-conversion}}$ is computed directly from the 10M sample. $C$ is picked through trial-and-error with $\{p_{\text{torso}},p_{\text{tail}}\}$ decided subsequently. $\lambda$ and $\alpha$ are estimated through maximum likelihood. An analytical solution for $\alpha$ exists (i.e. the Hill estimator \cite{Hill75}) and it indeed says that the tail has a shape between 1 and 2 meaning infinite variance. To see how well the truncated exponential and Pareto fit, Fig. \ref{fig2} (a) and (b) show their QQ plots. Distortion in the left suggests to drive down the approximation error by introducing an additional segment in-between, using e.g. log-normal. We choose not to do so in order to control the complexity. There is also deviation in the far right of Fig. \ref{fig2}(b) beyond \$10,000. It is not particularly worrisome since similar deviation can be found in plots against even true Pareto. We consider current models simple yet good enough for the purpose of this study. 

Imagine we sample from (\ref{eq_dis}) using two sets of $\lambda$ and $\alpha$, one for treatment and one for control. The winsorized estimation $\hat\mu_a$ is computed accordingly and compared against $\mu$, which can be derived from (\ref{eq_mean}). $a$ is to be explored within the range of $[95\%, 99.9\%]$ percentiles. We have three important notes to make which can be easily overlooked in practice causing potentially problematic experiments.

\begin{figure}
\centerline{\includegraphics[scale=0.4]{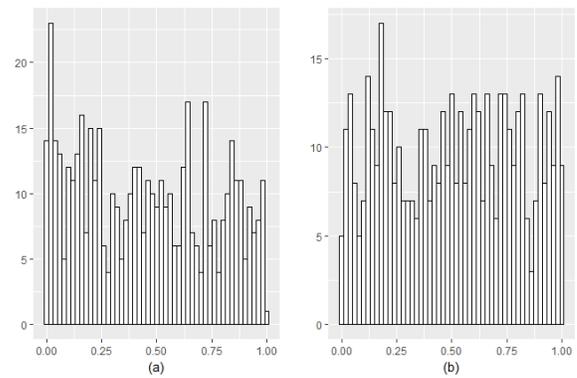}}
\caption{Histograms of p-value calculated from A/A tests. (a) Winsorization thresholds are calculated separately from and applied to treatment and control. (b) A unified threshold is calculated from the combination of both groups.}
\label{fig3}
\end{figure}

First, given treatment and control an immediate thought is to pick and apply their thresholds \emph{separately}, using e.g. the 99\% percentile. This will lead to unfortunately inflated significance, meaning higher than expected false positive rate. To illustrate we run 500 A/A tests, with treatment and control simulated from the same mixture model (or just sampled from the parent set). Fig. \ref{fig3}(a) shows the histogram of resulted p-value. We would expect to see a uniform distribution within [0, 1] and in particular, roughly 10\% falling into [0, 0.1]. Actually there are a lot more in this bucket, 79 out of 500. A binomial test reports 95\% confidence interval [12.7\%, 19.3\%] which clearly rejects a null hypothesis of 10\% false positive rate. Our suspicion is, as thresholds are selected using high percentile, they are volatile due to the wide spread in heavy tail and could differ notably between treatment and control. In fact, we easily find cases where the gap between two thresholds is over \$100, e.g. \$1069 vs. \$947 (both being 99\% percentile). It will cause the capped means of treatment and control more likely to have a big difference, thereby significant A/As. If this argument holds, we should be able to see significance diminish to expected level with \emph{unified winsorization} - picking the percentile from a combination of treatment and control then applying across both groups. Fig. \ref{fig3}(b) shows the histogram of p-value computed in this way. 46 out of 500 tests appear significant which is promising and a binomial test proves so with 95\% confidence interval [6.8\%, 12.1\%]. It is now a legitimate inference system exhibiting expected false positive rate.

\begin{figure}
\centerline{\includegraphics[scale=0.45]{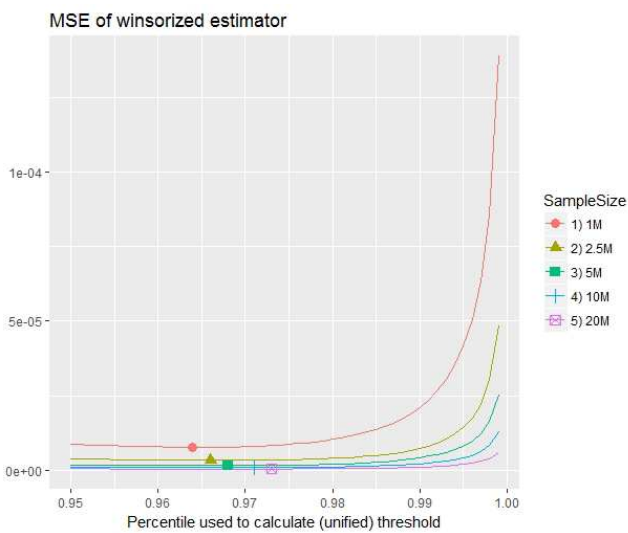}}
\caption{Selection of optimal percentile for winsorization differs upon sample size of an experiment.}
\label{fig4}
\end{figure}

Part of reason why the consequence of separate winsorization is murky is because it is less serious in experiments of moderate to large sample size, e.g. more than 2 million (assuming the same for treatment and control). In comparison it is only 500k in the simulation of Fig. \ref{fig3}. Our second note is about this dependency. Fig. \ref{fig4} shows how the mapping between MSE and percentile selection changes across experiments of different size, all else held constant (unified winsorization is in use). Points of optimal percentile (minimum MSE) are marked in different colors/shapes. Overall the estimation error drops as sample size increases: the 1M-sized experiment produces the highest curve and the 20M-sized one produces the lowest. MSE skyrockets towards the right end suggesting winsorization works a lot better than naive estimation (equivalent to winsorization by 100\% percentile). At the same time optimal threshold is moving to the right and we can use gradually higher percentiles to achieve minimum MSE. The impact of heavy tail is seen decreasing in experiments of bigger traffic. 

\begin{figure}
\centerline{\includegraphics[scale=0.45]{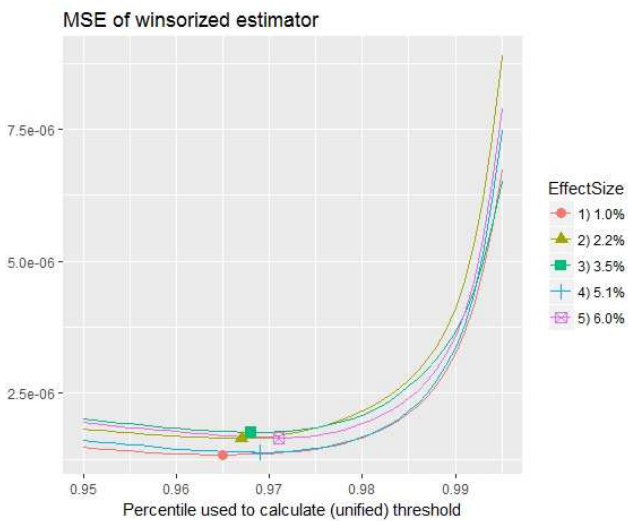}}
\caption{Selection of optimal percentile for winsorization differs upon the magnitude of treatment effect.}
\label{fig5}
\end{figure}

Sample size is only one of the complications for optimization. Our third note is regarding the magnitude of true treatment lift. In Fig. \ref{fig4} it is fixed at 1.5\% to simplify the study. Next we hold sample size at 5M and simulate instead a range of treatment lifts between 1\% and 6\%. Fig. \ref{fig5} depicts how the mapping between MSE and percentile selection changes accordingly. Again it is preferable to have winsorization deployed but the curves are more interleaved this time. Once the points of minimum MSE are marked, we are seeing the necessity to use gradually higher percentile along with more dramatic treatment effect. To a large extent this is expected because a bigger effect often implies more observations in the tail, a capping of which will be a loss of valid information. 

Given the dependencies a reasonable practice is to have a two-dimensional lookup table, where the optimal percentiles are stored based on sample size and true treatment lift. The table can be pre-computed using grid search. Sample size is specific to an experiment and usually determined through power analysis (with business constraints). However, foreseeable effect size runs the risk of being arbitrarily chosen. At eBay we have a rich history of experiment data so the proposal is to anchor it with a range of lifts learned from the past, e.g. [0.5\%, 1.5\%]. Experiment owners can still adjust but is bounded within the range. Since they also specify an expected lift in power analysis we can leverage the same tool and value. Ideally, the parameter can be determined together with sample size during experiment setup.

The notes we make here provide high level guidelines rather than implementation specs. In fact unified winsorization can be approached in different ways, e.g. simply using an average of the two separate percentiles rather than taking a percentile from the treatment and control union. The mixture model and MSE-percentile curves also vary across metrics. Following the guidelines we have a workable but cumbersome system - a large number of grid searches need be conducted in advance to build the lookup tables. A more approachable solution is therefore in demand. Next we will tap into the competency of robust statistics and have a proposal.

\section{Robust regression}
\label{section3}

Winsorization is a simple tweak to sample mean by putting a cap on observations. It is not surprising to see drastic reduction of variance in Fig. \ref{fig4} and \ref{fig5} since the contribution from observations beyond threshold is restrained. However, these `outliers' are not really incidental errors but genuine to the underlying, albeit heavy-tailed, distribution. Imagine one customer spends around a thousand dollar and another spends over a million but the two are weighted same. Even if we optimize the threshold it is arguably an awkward setting. Besides, sample mean can be derived as a maximum likelihood estimator where the observation has a Gaussian distribution, resulting in a quadratic loss $L_2=\epsilon^2$ where $\epsilon$ is the residual. By design it has the tendency to be dominated by extreme values. Relative efficiency of the estimator is going to be poor for a heavy-tailed distribution \cite{Maronna19}. 

We don't want to fall back to a solution requiring parametric models as in section \ref{section2}. The other way is to find estimators which could be robust against the tail behavior. Median is a popular choice which can be derived from an absolute loss function $L_1=|\epsilon|$. But it is biased with regard to estimating the mean when $\epsilon$ is not symmetric. If we visualize a better design, the loss function would have strong convexity in a uniform neighborhood of its minimum ($\epsilon=0$) but extend smoothly to an affine function at the boundary. These properties help combine much of the efficiency of quadratic loss and the robustness of absolute loss. Huber loss function is such a design \cite{Huber64}. Mathematically,
\begin{equation}
L_\delta(\epsilon)=\left\{
		\begin{array}{ll}
		\frac{1}{2}\epsilon^2, & \text{for }|\epsilon|\leq\delta\\
		\delta(|\epsilon|-\frac{1}{2}\delta), & \text{otherwise}
		\end{array}
	\right.
\end{equation}
where $\delta$ controls the width of neighborhood hence the blending of $L_2$ and $L_1$ losses. It is known as the tuning constant and set by default at 1.345 times standard deviation of $Y_i$ to achieve 95\% asymptotic efficiency. A notable advantage is its adaptivity across different metrics. We will follow this recommendation but keep an eye on its performance.

Huber is only one type of M-estimators which get the name by generalizing maximum likelihood estimation. Others include Tukey's biweight loss and Hampel loss that have redescending derivatives beyond the neighborhood. There are also L-, S-estimators and theories around breakdown point, influence function, etc. It is not our purpose to have a review of robust statistics in this paper and interested readers are referred to \cite{Maronna19}. We remain focused on introducing the technique, in particular Huber loss, to online experimentation and studying its behavior against heavy-tailed metrics.

\begin{figure}
\centerline{\includegraphics[scale=0.41]{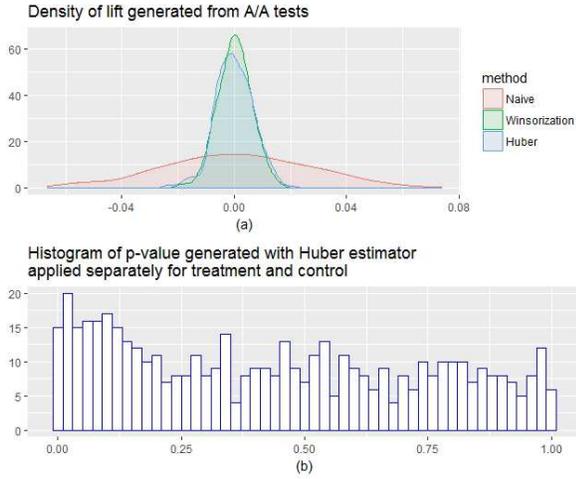}}
\caption{Direct application of Huber estimator.}
\label{fig6}
\end{figure}

A straightforward idea is to run Huber estimation for treatment and control separately and proceed with calculating the lift and p-value. Fig. \ref{fig6}(a) shows the distributions of resulted lifts from 500 A/A simulations (treatment/control is sized at 2M and sampled from the parent set). It is clear that both Huber and winsorization enjoy much less variance than the naive estimation. Then we plot the histogram of p-value generated from Huber in Fig. \ref{fig6}(b). Unfortunately it suggests a skew towards the low end. 93 out of 500 tests have p-value less than 0.1 and the 95\% confidence interval of false positive rate is [15.3\%, 22.3\%]. It is not a valid method and significance is inflated, as we have seen in separate winsorization.

\begin{figure}
\centerline{\includegraphics[scale=0.41]{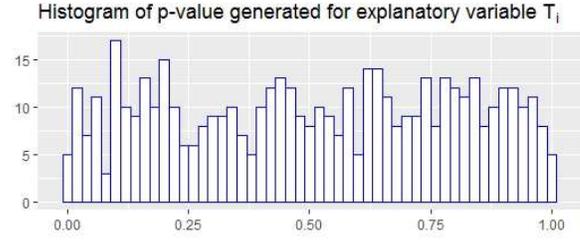}}
\caption{Huber regression applied in analyzing A/A test.}
\label{fig7}
\end{figure}

It is possible that some form of unified robustification could be a fix. For that we embrace a new framework in analyzing the experiment response. Consider a linear regression 
\begin{equation} \label{lr}
Y_{i}=\alpha+\beta T_{i}+\epsilon_i
\end{equation}
over the pair $(T_i, Y_i)$ where the explanatory variable $T_i$ labels a customer's assignment (0 for control and 1 for treatment). $\alpha$ is the intercept. A least square fitting of coefficient $\beta$ is in fact equivalent to the naive estimation of treatment effect \cite{Gelman06}. z-test of $\hat\mu=0$ also turns out same as testing the significance of $T_i$. There are more advantages to this framework but for now Huber loss plays nicely in replacing least square, which is not robust in the presence of heavy tail. Training the model is essentially minimizing one join loss function rather than two separate for treatment and control. If it works, we should be able to see the significance of $T_i$ generated from A/A tests distribute uniformly within [0, 1]. The result shown in Fig. \ref{fig7} corroborates this notion. The data used is the same as in Fig. \ref{fig6} but 45 out of 500 tests report p-value less than 0.1, well aligned with our expectation. Legitimacy of the proposal is therefore established.

Next is to study the performance of Huber regression in comparison to naive estimation and winsorization. A visual contrast like in Fig. \ref{fig6}(a) is illuminating but imprecise. Numerical reports will be presented instead. Because reduction of variance often comes with a price of bias increase, besides MSE we also measure the mean absolute deviation (MAD), defined as $\mathbb{E}(|\hat\mu-\mu|)$. Together the two provide a balanced view. We will explore three configurations of sample size, i.e. \{2.5M, 5M, 10M\}, coupled with three values of expected lift, i.e. $\mu\in\{0.5\%, 1\%, 2.2\%\}$. Each combination will carry 500 simulated tests, generating treatment and control samples from (\ref{eq_dis}). The analysis methods are then applied, among which winsorization has its percentiles selected as before.

Table \ref{tab1} reports the result. Given the random nature it is risky to over read into differences within e.g. one order of magnitude. Still, we have the following comments:
\begin{itemize}
\item Both winsorization and Huber regression deliver much lower MSE across all combinations. The improvement over naive estimation is around three orders of magnitude.
\item Between winsorization and Huber the difference of MSE is within half order, which is fairly small. Winsorization is better in five cases while Huber wins in the other four.
\item The three methods are largely close in terms of MAD but winsorization does appear consistently worse. Naive estimation and Huber win in four and five of the cases respectively.
\end{itemize}
\begin{table*}
\caption{Comparison of MSE and MAD between naive estimation, unified winsorization and Huber regression. Minimum among the three methods is highlighted in bold.}
\begin{center}
\begin{tabular}{|l|l||l|l||l|l||l|l|}
\hline
\multicolumn{2}{|l||}{}                        & \multicolumn{2}{l||}{$\mu=0.5\%$} & \multicolumn{2}{l||}{$\mu=1.0\%$} & \multicolumn{2}{l|}{$\mu=2.2\%$} \\ \hline
Sample size           & Method                & MSE         & MAD        & MSE       & MAD       & MSE       & MAD       \\ \hhline{|=|=||=|=||=|=||=|=|}
\multirow{3}{*}{2.5M} & Naive estimation      & 5.58e-03	& -4.71e-03  & 1.91e-02  & 1.68e-02  & 6.62e-03  & -4.19e-03 \\ \cline{2-8} 
                      & Unified winsorization & 6.47e-06    & -2.21e-03  & \textbf{3.34e-06}  & -4.27e-03 & \textbf{3.03e-06}  & -7.06e-03 \\ \cline{2-8} 
                      & Huber regression      & \textbf{3.52e-06}    & \textbf{-1.61e-03}  & 6.25e-06  & \textbf{-2.71e-03} & 5.68e-06  & \textbf{-3.12e-03} \\ \hline
\multirow{3}{*}{5M}   & Naive estimation      & 3.41e-03    & \textbf{4.44e-04}   & 4.70e-03  & \textbf{2.20e-03}  & 2.87e-03  & \textbf{1.30e-03}  \\ \cline{2-8} 
                      & Unified winsorization & \textbf{1.66e-06}    & -2.12e-03  & 3.23e-06  & -4.25e-03 & 2.77e-06  & -6.96e-3  \\ \cline{2-8} 
                      & Huber regression      & 3.01e-06    & -1.36e-03  & \textbf{1.75e-06}  & -2.69e-03 & \textbf{1.35e-06}  & -2.89e-03 \\ \hline
\multirow{3}{*}{10M}  & Naive estimation      & 2.44e-02    & 7.55e-03   & 1.22e-03  & 3.32e-03  & 1.49e-03  & \textbf{1.39e-03}  \\ \cline{2-8} 
                      & Unified winsorization & 1.33e-06    & -1.98e-03  & \textbf{8.40e-07}  & -4.41e-03 & \textbf{7.92e-07}  & -7.02e-03 \\ \cline{2-8} 
                      & Huber regression      & \textbf{8.15e-07}    & \textbf{-1.21e-03}  & 1.65e-06  & \textbf{-2.89e-03} & 1.44e-06  & -2.88e-03 \\ \hline
\end{tabular}
\label{tab1}
\end{center}
\end{table*}

\begin{figure}
\centerline{\includegraphics[scale=0.44]{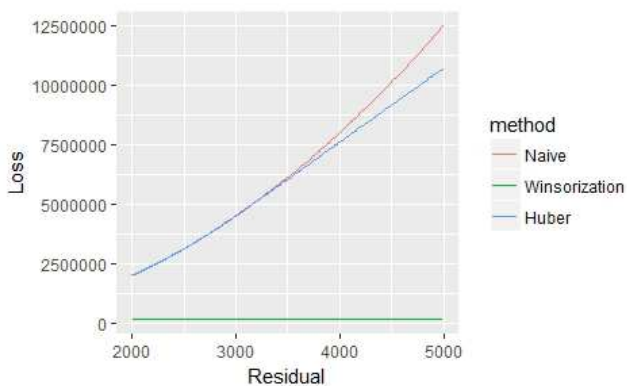}}
\caption{Comparison of loss function between naive estimation, unified winsorization and Huber regression.}
\label{fig8}
\end{figure}

High MSE leads to a wide confidence interval and thereby low detectability of an experimentation system. We have just seen Huber regression greatly improves that upon naive estimation, which is very good news to product teams. As for bias, no obvious degradation is observed. The benefit is known to build upon a redesign of loss function so it helps to take a look at how it changes on the ground. Fig. \ref{fig8} compares the three methods for residuals within [2000, 5000]. We zoom into the range because the curves would overlap in the neighborhood of 0 yet in the far end visualization gets unwieldy. Huber starts to deviate from naive estimation at around $\epsilon=3400$, featuring a linear instead of quadratic growth of loss. Impact of the tail observations upon estimation is thereby restrained. Winsorization in contrast goes too far by keeping the loss flat regardless of the growth of residual. When it is coming from observations genuine to the distribution, bias is injected. Huber regression in our view strikes an attractive balance.

Another advantage of this method is its simplicity for implementation. We avoid the complexity of building a large number of lookup tables for optimal winsorization. The default tuning constant is driven by observed data on the fly and performs fairly well. There are also off the shelf, easy to use software libraries. A price paid is of course the computation (no closed form solution exists so an iterative approach is required), but that proves to be affordable in today's environment. On the other side we have seen methods in the literature designed for asymmetric residuals. Although they could be more appropriate, building e.g. a number of quantile regressions \cite{Kanamori06} is not preferred in practice.

\section{Debiased machine learning}
\label{section4}

So far our answers to the estimation challenge, whether winsorization or Huber regression, address the heavy tail of response metric directly. There is another thread which works by incorporating additional information to account for part of the variance. In the last section we embark on a regression framework for the merit of unified robustification. Luckily it also gives us the flexibility to bring in more explanatory variables. Let $\mathbf{X_i}$ be a vector of them, known as the covariates. We then have an extended model
\begin{equation}
\label{eq_lm}
Y_{i}=\alpha+\beta T_{i}+\mathbf{X_i}\mathbf{\theta}^\intercal+\epsilon_i
\end{equation}
where $\beta$ again estimates the treatment efffect. Deng et al. reach the same in \cite{Deng13} but come from a different perspective, drawing analogy to variance reduction in Monte Carlo sampling. Their proposal is to use covariates pulled from pre-experiment period which are guaranteed to be independent of treatment assignment. It is not constructing strata for sampling to follow but retrospectively mining them after data collection. For this reason the method is often called \textit{post-stratification}. In order to be effective it need choose covariates having good correlation with response metric. The importance is proved mathematically and also in practice.

Eq. (\ref{eq_lm}) forces a linear relationship between $\mathbf{X}$ and $Y$ which is debatable. When we have millions of data points and potentially high dimensionality in covariates, as in online experiments, it is reasonable to ask if nonlinear and more sophisticated functions could serve better. Imagine a model, 
\begin{equation}
\label{eq_nlm}
Y_{i}=\beta T_{i}+f(\mathbf{X_i})
\end{equation}
where $f(.)$ can take forms not just in polynomials but also those convoluted ones learned using modern algorithms (residual is skipped). We would hope the enhancement in expressiveness can help drive more variance reduction in $\beta$. Note $T_i$ is additively separable, so this model continues to aim for estimating ATE. There is a way to accommodate general heterogeneity which is to have $Y_{i}=f(T_i, \mathbf{X_i})$ instead. It is interesting for future work but beyond the scope of this paper. 

Traditionally, much of experimentation practice emphasizes on treatment effect and leaves the fitness of $f(.)$ unrecognized \cite{Athey15}. Researchers often specify a model, train it on the data, and rely on statistical theory to build confidence intervals for $\beta$. They may hand-pick more specifications behind the scene to check for robustness of estimation but rarely report it. Concerns are that multiple testing could invalidate confidence intervals and even the lure of fishing for models with desired results. We believe in the value of systematic and disciplined model exploration and optimization, something that is more prominent in the big data era with assistance of machine learning techniques. But the exploitation is not straightforward and ought to be handled with great care.

To build some intuition how ML is different from causal inference, it is useful to think of the following case. Suppose we have a data set containing customer spending and claim. If the goal is to have an accurate estimate of spending, we shall find increased number of claim is predictive of higher spending since customers will be paying more attention. What is learned from the data is really correlation between claim and spending. In contrast, imagine we wish to know how the spending would change if we make the submission of claim easier. This is a question of causal inference. Clearly we won't conclude that getting customers claim more will drive the spending up. It is well known that correlation does not imply causation. The latter cannot be answered simply by examining history without additional assumptions or structure. If the data is generated from controlled experiments, it can be safely used to draw a conclusion; otherwise we will turn to observational research or instrumental variables. Interestingly when we project the covariates onto selected instruments, only part of their variation is utilized in estimating response metric. The predictive power is expected to be lower. In other words we deliberately abandon the goal of accurate prediction in pursuit of an unbiased estimate of causal parameter.

Here we have a controlled setting so the identifiability of causal effect is not at question. But it is far from error-proof. A central theme in ML is the tradeoff between model expressiveness and risk of overfitting. When a model has too many adjustable terms relative to sample size, the goodness of fit measured out of sample on an independent test set is expected to be much worse than measured on the sample. A popular cure is cross-validation where we partition the sample into complementary subsets, repeatedly estimate a model on some of them (``training'') and evaluate it on the rest (``test''). Model selection is by minimizing e.g. average prediction error over the test folds. We probably end up with models of reduced complexity, manifested as reduced degree in polynomials, decision tree or network pruning, or omitted covariates in LASSO. Because optimization is not targeting the accuracy of parameters, these models can be very misleading. Imagine the relationship between omitted covariates and response is loaded onto the causal parameter. We may have an easily comprehensible model but the information gained is incomplete or even wrong.

There is more discussion of the nuances in \cite{Athey18}. To help harness the power of ML while maintaining the integrity of causal inference, a number of techniques have been developed. Before looking into them it makes sense to review \textit{propensity score matching} (PSM) first, a method towards modeling the relationship between covariates and the treatment assignment:
\begin{equation}
\label{eq_psm}
T_{i}=g(\mathbf{X_i})
\end{equation}
where residual is skipped. Since $T_i$ is binary a logistic regression or classification algorithm is well suited. Treatment effect is measured as before but each $Y_i$ is now given a weight
\begin{equation}
w_{i}=\frac{T_i}{g(\mathbf{X_i})}+\frac{1-T_i}{1-g(\mathbf{X_i})}.
\end{equation}
The method is critical to observational research where the assignments could be confounded by covariates. In a controlled setting we are interested instead in its role in variance reduction.

We won't know whether any particular model, $f(.)$ or $g(.)$, accurately reflects the respective relationship or not. Attention of developments is hence on the robustness against potential misspecifications. The idea of \textit{doubly robust} \cite{Robins95} is to combine the response model with assignment model so that we can have unbiased estimation even with one of the two misspecified. Let $\tau(T_i, f(\mathbf{X_i}))$ be the prediction of response from (\ref{eq_nlm}). Treatment effect is calculated as
\begin{multline}
\label{eq_dr}
\mathop{\mathbb{E}}[\tau(1,f(\mathbf{X_i}))-\tau(0,f(\mathbf{X_i}))] \\
+\mathop{\mathbb{E}}[\frac{T_i}{g(\mathbf{X_i})}(Y_i-\tau(1,f(\mathbf{X_i})))+\frac{1-T_i}{1-g(\mathbf{X_i})}(Y_i-\tau(0,f(\mathbf{X_i})))].
\end{multline}
It is an additon of two expectations and PSM is only applied to the residuals from response model. 

The challenge of covariate selection is to control model complexity and at the same time retain important regressors. For that Belloni et al. \cite{Belloni14} propose a three-step procedure: first use LASSO to select variables that predict the assignment, then use LASSO to select variables that predict the response (ignoring assignment), finally take a union of selected variables and run the regression (\ref{eq_lm}) by ordinary least squares. The method is seen very promising in high dimensional applications.

Recently Chernozhukov et al. \cite{Cherno17} show that (\ref{eq_dr}) can be recast in terms of Neyman orthogonality conditions which are robust to perturbations in functions $f(.)$ and $g(.)$. With this advancement in theory their proposal is to have \textit{double/debiased} machine learning. For example, use only half of data to run two separate learning tasks, one for $f \colon \mathbf{X} \to Y$ and one for $g \colon T \to Y$; predict over the other half and take the residuals 
\begin{equation}
\epsilon^Y_i=Y_i-f(\mathbf{X_i}) \quad\text{and}\quad \epsilon^T_i=T_i-g(\mathbf{X_i}), 
\end{equation}
then linearly regress 
\begin{equation}
\label{eq_dml}
\epsilon^Y_i=\beta\epsilon^T_i
\end{equation}
to have an estimate of $\beta$. Swap the two halfs to have another estimate and average the two to produce the final result. Cross-fitting splits the data into complementary subsets, one of which is used for model learning and the other for parameter estimation. It is known useful to attack potential overfitting with complex models. In the mean time the orthogonal design protects against general misspecifications. The above procedure is proven to remove estimation bias asymptotically. 

\subsection{Empirical results}
We want to bring some of these developments into online experimentation and verify their performance. Motivation is of course to reduce estimation variance while controlling bias with heavy-tailed metrics. Like in \cite{Deng13} we pull covariates from the pre-experiment period. While a variety of learning algorithms can be used to construct $f(.)$ and/or $g(.)$, we are particularly interested in Huber regression studied in section \ref{section3}. One difference from previous is the data set - we won't use the customer spending sample here. The reason is that a lot of customers are absent to site in e.g. the two weeks period right before experiment, causing null values in covariates. There is no way to backfill or simulate them with accuracy. 

At eBay besides sampling by customer we also support other strategies, one of which is by merchandise category (e.g. ``DVDs \& Movies:DVDs \& Blu-ray Discs'') and provides the desired stability across pre and post periods. It has been used in a price guidance experiment to avoid potential interference between treatment and control \cite{Wang18}. Success metric is per category sales in dollar amount which is again heavy-tailed. Sample size is yet much smaller given total about 10k leaf categories. Detectability of a naive estimation would be disappointing without the help of covariates. So for every category we not only have the in-experiment sales but also following regressors, all from the pre-experiment period:
\begin{itemize}
\item sales of the same category
\item number of sold items
\item number of listed items
\item average item price
\item ratio of relisted items
\item ratio of new items
\item ratio of multi-SKU items
\item ratio of fix-priced items
\item ratio of free-shipping items
\item ratio of items having product identifiers (MPN/UPC/GTIN/etc.)
\item ratio of items with returns accepted
\end{itemize}
The list is largely out of business considerations. It is short relative to sample size so we won't pursue LASSO or Belloni's method \cite{Belloni14}. These covariates, especially the first one, will be seen playing a big role in variance reduction. 

The study is based on 500 A/A simulations where we know the ground truth $\mu=0\%$ and can compute MSE and MAD. Nine methods are compared accordingly. Besides the three which don't leverage covariates (naive estimation, unified winsorization and Huber regression), we implement 
\begin{itemize}
\item post-stratification
\item post-stratification with Huber regression used in (\ref{eq_lm})
\item PSM; a logistic regression is used
\item doubly robust
\item double/debiased learning (five-fold cross-fitting)
\item double/debiased learning with Huber regression used in (\ref{eq_dml}); other configurations are kept same
\end{itemize}
The results are reported in table \ref{tab2}. We refrain from testing other models in $f(.)$/$g(.)$ not to dilute the focus.
\begin{table}
\caption{Comparison of MSE and MAD across nine methods used to estimate treatment effect from A/A tests. Minimum is highlighted in bold.}
\begin{center}
\begin{tabular}{|l||l|l|}
\hline
Method                   & MSE & MAD \\ \hhline{|=||=|=|}
Naive estimation                  & 1.979e-02      & 1.012e-02    \\ \hline
Unified winsorization             & 7.586e-03      & 4.098e-03    \\ \hline
Huber regression                  & 1.792e-05      & 1.705e-04    \\ \hline
Post-stratification               & 3.559e-04      & 6.223e-04    \\ \hline
Post-stratification w/ Huber      & 2.887e-07      & 1.789e-05    \\ \hline
PSM                               & 1.434e-03      & 1.089e-03    \\ \hline
Doubly robust                     & 3.711e-04      & 4.689e-04    \\ \hline
Double/debiased learning          & 2.296e-05      & 1.683e-04    \\ \hline
Double/debiased learning w/ Huber & \textbf{1.453e-07}      & \textbf{-1.159e-05}   \\ \hline
\end{tabular}
\label{tab2}
\end{center}
\end{table}

Immediately we notice that double/debiased learning with Huber regression ranks best in both criteria. The only method which achieves close enough performance is post-stratification with Huber regression. Gap between the two is coming from propensity score matching. In the second tier there are double/debiased learning (vanilla version) and Huber regression. They make effort on two fundamentally different fronts, one leveraging covariates and the other attacking the heavy tail. The third tier contains post-stratification and doubly robust. Following are PSM, unified winsorization and lastly, naive estimation. The refinement of estimator is seen generating enormous value: reduction of variance is around five orders of magnitude and reduction of bias is around three orders. Speaking of detectability (approximated by two standard deviation), we are only capable to discover \textasciitilde28.2\% lift using the naive method and \textasciitilde17.4\% using winsorization, compared to \textasciitilde0.9\% by Huber regression, \textasciitilde0.1\% by post-stratification w/ Huber and \textasciitilde0.08\% by double/debiased learning w/ Huber. Bias of the system is actually diminished at the same time.

\begin{figure}
\centerline{\includegraphics[scale=0.4]{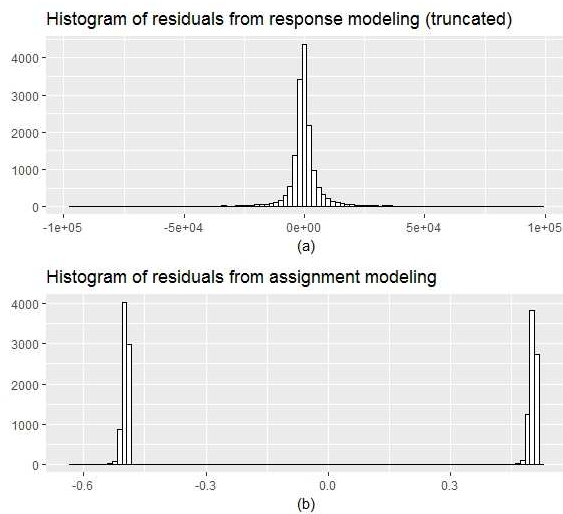}}
\caption{A look into the residuals from modeling experiment response (a) and assignment (b).}
\label{fig9}
\end{figure}

Double/debiased learning is proved to be a critical ingredient to the gain. We are not measuring treatment effect directly from $(T_i, Y_i)$ but from the residuals $(\epsilon^T_i, \epsilon^Y_i)$ instead. They are prediction errors of a response model and an assignment model which are learned \textit{orthogonally}. A visualization of the two is provided in Fig. \ref{fig9}. It is worth noting that despite a long tail (not shown here), $\epsilon^Y_i$ is fairly symmetric compared to $Y_i$, accommodating comfortably a symmetric loss function like Huber. The distribution of $\epsilon^T_i$ concentrates around +0.5 and -0.5, implying that our prediction of assignment is consistently close to 0.5. So we do have random sampling conditioned upon $\mathbf{X_i}$. There is sizable benefit from exploiting the response model alone (post-stratification); the assignment model (PSM) helps less since we operate in a controlled environment. Combination of the two is not straightforward. We don't see much of further gain by doubly robust. But the orthogonal design is beneficial in a significant way, corroborating the theoretical result in \cite{Cherno17}. Even better the integration with Huber loss is smooth, producing a tight system easy to implement and manage. Although steps like cross-fitting and running regressions for response/assignment add to the computational cost, it is well-grounded without a doubt.

\section{Conclusions and future work}
\label{section5}
Knowledge discovery from experiment data has a long history dating back to Sir Ronald Fisher's work in the early nineteenth century. Since then it has seen developments in increasingly broad applications. Over a decade into the new century, the proliferation of Internet poses a substantial challenge. Millions of observations, often much more, are readily available and people are anticipating highly accurate measurements. The observation may contain not just few directly concerning variates but a rich set of covariates. There is potentially a wealth of functional relationship to be modeled and exploited. On another side irregularities like heavy-tailed distributions are very common among the variates. Making causal inference under these circumstances is inevitably complicated. As product decisions are at stake, stringent control of both integrity and detectability is demanded in delivering experiment analyses.  

Multiple solutions are explored in this paper in a systematic manner. We start with the response metric and show that its heavy-tailed behavior is a fundamental difficulty to confront. Winsorization is simple and effective in driving down the estimation variance. However, it is loosely defined in our view and leads to inflated false positives if not properly designed. The necessity of unified winsorization is then discussed, as well as how it ought to be optimized according to sample size and foreseeable treatment effect. We are not satisfied with the design and turn to the theory of robust statistics for inspiration. Huber loss provides a more delicate way to handle extreme observations. Its derivative, Huber regression, is proved a legitimate and superior alternative. On leveraging covariates we tap into the latest development of debiased machine learning. While sophisticated models can be built towards predicting the response and assignment, a design of orthogonal composition together with cross-fitting is demonstrated rewarding. A compelling choice is then to integrate Huber regression inside the framework, bringing forth a reliable inference system for decision making. We also take an extensive usage of simulation techniques and A/A tests, proven to be instrumental in this journey. 

There are a number of topics which we have in consideration for future work. To mention some,
\begin{itemize}
\item It is worthy testing advanced models like random forest or neural network in contructing $f(.)$ and $g(.)$. We also notice the development of \textit{causal forest} and \textit{honest} estimation in \cite{Wager18}. Closely related is to build confidence interval for the point estimate, which is largely neglected so far. Chernozhukov et al. \cite{Cherno17} also suggest an approach which takes into account the uncertainty due to sample splitting. The field is growing very fast and deserves a close watch.
\item Catoni's breakthrough \cite{Catoni12} reveals that for heavy-tailed distribution the empirical risk calculated upon training data is no longer a good approximation. It inspires recent developments on learning with heavy-tailed losses, including \cite{Zhang18} which proposes a truncated minimization approach for $L_1$-regression. It will be interesting to see how some of these could benefit our application. 
\item Many practitioners would resort to transformations like log-normal to tackle the data skewness. But the transformed data may not exhibit a desired distribution. More importantly, the resulted parameter estimate is often biased and hard to interpret. The smearing retransformation \cite{Duan83} could be an answer but its merits and potential intricacy need be examined.
\end{itemize}

\begin{acks}
  David Goldberg helped a lot when he was with eBay, especially in the design of winsorization. The authors would also thank rest of eBay Experimentation Science team for their suggestions. 
\end{acks}

\bibliographystyle{ACM-Reference-Format}
\bibliography{ms_bib}

\end{document}